\documentclass[acmsmall]{acmart}

\AtBeginDocument{%
  \providecommand\BibTeX{{%
    \normalfont B\kern-0.5em{\scshape i\kern-0.25em b}\kern-0.8em\TeX}}}

\copyrightyear{2021}
\acmYear{2021} 
\setcopyright{rightsretained}
\acmConference[CHI '21 Extended Abstracts]{CHI Conference on Human
Factors in Computing Systems Extended Abstracts}{May 8--13,
2021}{Yokohama, Japan}
\acmBooktitle{CHI Conference on Human Factors in Computing Systems
Extended Abstracts (CHI '21 Extended Abstracts), May 8--13, 2021,
Yokohama, Japan}\acmDOI{10.1145/3411763.3443432}
\acmISBN{978-1-4503-8095-9/21/05}

\begin{document}

\title{VR Hackathon with Goethe Institute: Lessons Learned from Organizing a Transdisciplinary VR Hackathon}

\author{Wiesław Kopeć}

\email{kopec@pja.edu.pl}

\affiliation{%
  \institution{Polish-Japanese Academy of Information Technology}
}

\author{Krzysztof Kalinowski}
\affiliation{%
  \institution{Polish-Japanese Academy of Information Technology}
}

\author{Monika Kornacka}
\affiliation{%
  \institution{SWPS University of Social Sciences and Humanities}}

\author{Kinga H. Skorupska }
\affiliation{%
  \institution{Polish-Japanese Academy of Information Technology}}

\author{Julia Paluch}
\affiliation{%
  \institution{Polish-Japanese Academy of Information Technology}}

\author{Anna Jaskulska}
\affiliation{%
  \institution{KOBO Association}}

\author{Grzegorz Pochwatko}
\affiliation{%
  \institution{Institute of Psychology Polish Academy of Sciences}}

\author{Jakub Filip Mozaryn}
\affiliation{%
  \institution{Warsaw University of Technology}}

\author{Paweł Kobyliński}
\affiliation{%
  \institution{National Information Processing Institute}}

\author{Piotr Gago}
\affiliation{%
  \institution{Polish-Japanese Academy of Information Technology}}

\renewcommand{\shortauthors}{Kopeć et al.}
\renewcommand{\shorttitle}{VR Hackathon with Goethe Institute: Lessons Learned}
\begin{abstract}
  In this article we report a case study of a Language Learning Bauhaus VR hackathon with Goethe Institute. It was organized as an educational and research project to tap into the dynamics of transdisciplinary teams challenged with a specific requirement. In our case, it was to build a Bauhaus-themed German Language Learning VR App. We constructed this experiment to simulate how representatives of different disciplines may work together towards a very specific purpose under time pressure. So, each participating team consisted of members of various expert-fields: software development (Unity or Unreal), design, psychology and linguistics. The results of this study cast light on the recommended cycle of design thinking and customer-centered design in VR. Especially in interdisciplinary rapid prototyping conditions, where stakeholders initially do not share competences. They also showcase educational benefits of working in transdisciplinary environments. This study, combined with our previous work on human factors in rapid software development and co-design, including hackathon dynamics, allowed us to formulate recommendations for organizing content creation VR hackathons for specific purposes. We also provide guidelines on how to prepare the participants to work in rapid prototyping VR environments and benefit from such experiences in the long term.
\end{abstract}

\ccsdesc[500]{Human-centered computing~Virtual reality}
\ccsdesc[300]{Human-centered computing~HCI design and evaluation methods}
\ccsdesc[300]{Human-centered computing~Empirical studies in HCI}

\keywords{hackathon, language learning, teamwork, VR, participatory design}

\begin{teaserfigure}
  \includegraphics[width=\textwidth]{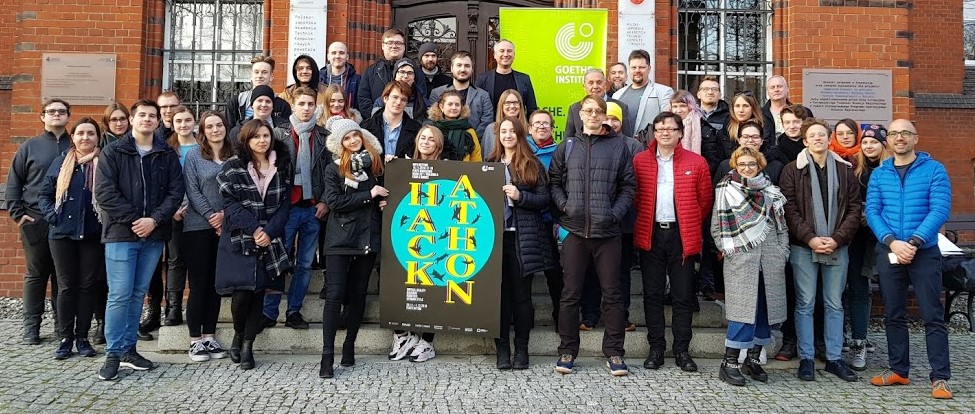}
  \caption{Participants and team at the PJAIT Innovative Technologies Research Centre in Bytom, Poland}
  \Description{Some of the hackathon participants and staff in front of our venue: The PJAIT Innovative Technologies Research Centre in Bytom}
  \label{fig:teaser}
\end{teaserfigure}

\maketitle

\section{Introduction} 

Team collaboration of experts in multiple fields is necessary to break new ground in multiple areas of industry and research while working at the intersection of ICT, HCI and other disciplines. Yet, engaging a team of experts in projects for specific purposes tends to be resource-intensive and time-consuming. Even more so if the projects involve VR, where low-budget prototyping is quite difficult and lengthy,
especially when co-designing with stakeholders who have little prior experience with VR. For this reason, it is important to explore collaboration schemes which not only cut the cots and allow for low fidelity VR prototyping of content to increase creativity, but also engage all stakeholders in the design process. This need is closer to "design for user empowerment", as proposed by Ladner \cite{ladner2015design} than the traditional "human centered design", which uses tools approximating the needs of the real target group rather than direct involvement of stakeholders in the design process.
To see how representatives of different disciplines may work together towards a very specific purpose under time pressure in the context of VR, we have organized an educational and research project in the form of a VR hackathon. This study allowed us to evaluate our approach to VR hackathons as a rapid prototyping tool for building VR solutions for specific purposes. It also showcased the dynamics of work in transdisciplinary teams consisting of stakeholders representing different disciplines, in our case Linguistics, Design, Psychology and VR tech. What follows is a case study of a Language Learning Bauhaus VR hackathon with Goethe Institute where the interdisciplinary teams were challenged with a specific requirement: to build a Bauhaus-themed German Language Learning VR Application. 

\begin{figure}[tb]
 \centering 
 \includegraphics[width=\columnwidth]{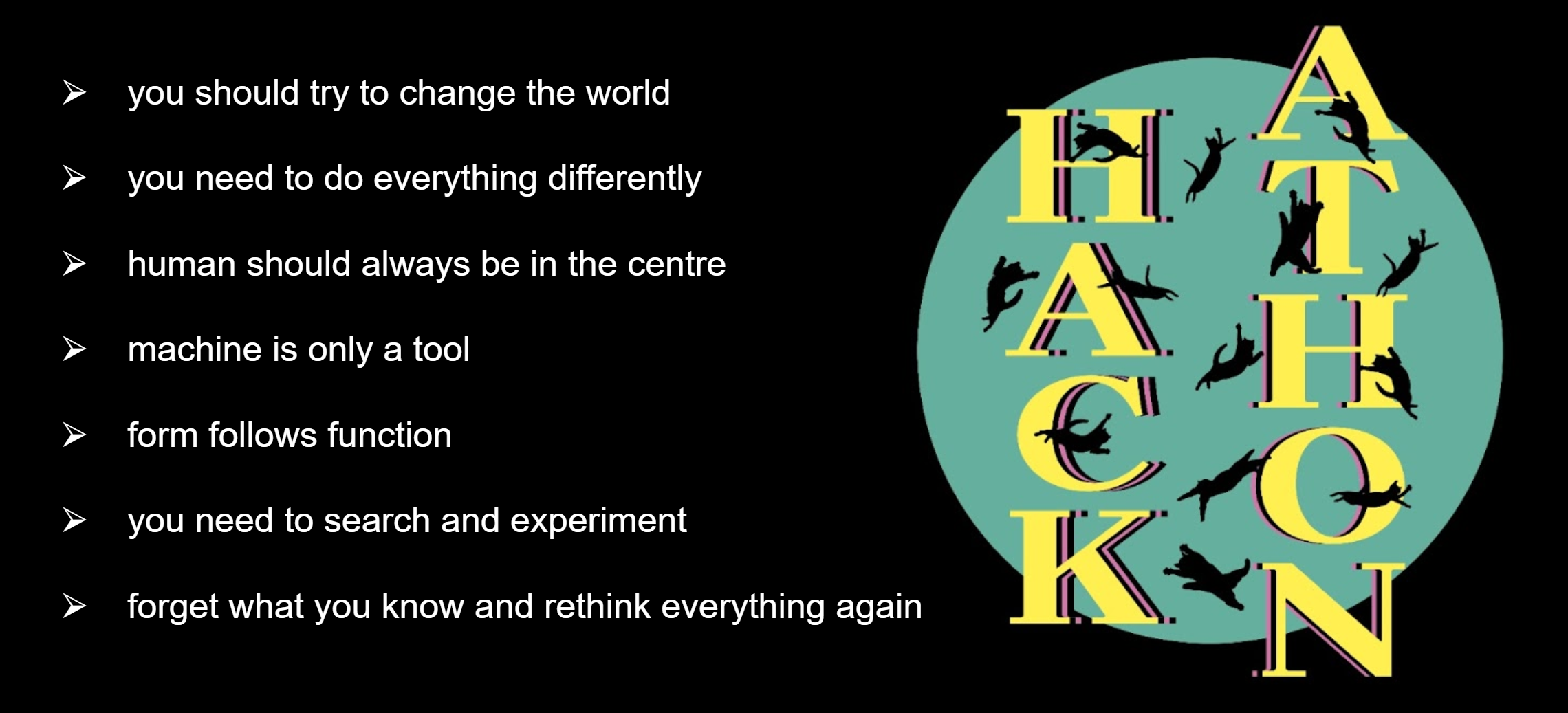}
 \caption{A summary of key points related to the theme of the VR Hackathon - the Bauhaus school of art}
 \Description{A summary of key points related to the theme of the VR Hackathon - the Bauhaus school of art}
 \label{fig:baukeys}
\end{figure}

\section{Literature Review}

Hackathons are hacking marathons, usually held over 24 hours, where individuals or teams work towards a solution to a presented problem, usually in the field of software engineering. 
There are marked benefits of hackathon participation for students \cite{nandihackathons}, especially in STEM disciplines and when it comes to informal learning \cite{fowler2016informal}. This format is increasingly popular and more widespread, with multiple researchers gathering best practices \cite{lapp20072006} and insights, for example, into the long-term effects of hackathons on innovation \cite{dehlihackathons}. Their variety can be seen in a classification developed by Briscoe in 2014 \cite{briscoe2014digital} that takes into account the themes and focus of different types of hackathons. The same review points out how women are often underrepresented in hackathons, but those teams which include them, or other diverse groups suffer from lower performance \cite{elsass1997demographic} possibly owing to time constraints and unconscious biases due to snap judgments of categorizing individuals in the team. To mitigate this problem Flippova et al. \cite{filippova2017diversity} suggests brainstorming at the start of the event, and Kopeć et al. suggests engaging in the empowerment stage \cite{kopec2017spiral}. Overall, hackathons are very valuable learning experiences, which may inspire creators to adapt multiple areas of civil society (education, government or market economy) to this new reality, as noticed by Prieto et al.\cite{mit2019hackathon} in their account of taking part in an MIT VR Hackathon.

\section{VR Hackathon}

 We constructed this experiment as a VR Hackathon case study, a very effective form of exploratory qualitative research \cite{rossman2003learning}, which is particularly effective to study phenomena which are rather difficult to be examined separately from the context, as postulated by Yin \cite{yin2013case}. We used this approach to simulate how representatives of different disciplines may work together towards a very specific purpose under time pressure in rapid prototyping conditions in the context of VR. 

\subsection{Key information}

\subsubsection{The Specific Requirement}

The goal of the hackathon was to prototype a Bauhaus-themed German Language Learning VR App. 

The target group of the app were to be young adults, whose linguistic competence is around A2-B1 according to Common European Reference Levels and who are open to new experiences in VR, as well as curious about the German culture. This was done in the context of a bigger goal to make learning German more attractive to young people who may want to study or work in the German language. 
Goethe Institute seeks to intertwine German learning with Virtual and Augmented reality by engaging with tech-savvy partners and forming innovative cooperation schemes. These could pave the way for more modern and engaging ways of learning the German language.

\subsubsection{Theme}

The theme was connected to the celebration of the 100th founding anniversary of the Bauhaus in 2019. \footnote{The centenary celebrations were organized under the motto "Thinking the World Anew" which resonates very well with VR. More on The Bauhaus can be found here: https://www.bauhaus100.com/the-bauhaus/ } Based on the lectures by Michael Fleischer some key points were given to clarify the ideas behind this movement can be seen in Figure \ref{fig:baukeys}. In short, the Bauhaus was a German art school of fine arts and crafts whose ideas heavily influenced modern art and design practice and education. Bauhaus was also chosen because of its approach to experimenting in education, open-mindedness to new methods and the idea that form follows function; all of these ideas were well-aligned with the goal of this VR hackathon.

\subsubsection{Research Tools}

We have invited multiple experts to act not only as mentors and members of the jury, but also to observe the dynamics of the team formation, collaboration, work and the quality of the final products. Therefore, the key tool in this case, as most of our data is qualitative, was participatory observation and analysis.

\subsection{Organization}

\subsubsection{Recruitment and Timeline}
The Acknowledgments section describes how cooperation was initiated, as it warrants special thanks. Proper preparations for the Hackathon started with a joint recruitment scheme: students of different faculties from several universities with diverse profiles (technology and social sciences) were invited to apply through an application form at the Goethe Institute website. The recruitment call was followed by a kick-off meeting, which explained the objective, theme and principles of the hackathon, presented attractions (such as the tour of the Hackathon venue and access to the variety of top-notch solutions across the VR HMD board) and announced the awards: the possibility to present the outcomes to the greater audience at a special event and an all-inclusive trip to Berlin to present the winning VR app.

The organizational timeline of the hackathon and related events was as follows: 
\begin{enumerate}
\item The kick-off meeting: 15.11.2019 
\item The hackathon: 29.11. – 1.12.2019 
\item Follow-up meetings and consultations: scheduled after the hackathon and before forthcoming presentations (Winter 2019/2020)
\item Presentation of winners 1–3 in the context of the Themenraum programme: March 2020 
\item Participation of winners in an international VR-Training at Goethe-Institut Berlin: April 2020
\end{enumerate}

\subsubsection{Venue, Technology and Supplies}

Our venue was an Innovative Technologies Research Centre 
which houses the Human Motion Lab, the Human Seeing Lab, The Human Facial Modelling Lab as well as Human Microexpression Lab and Wearable Technologies Lab. Part of the appeal of the hackathon was the possibility to get a quick tour of the labs on day 1, before the start of the hackathon.
During the core part of the contest, 2 of 5 VR kits were situated in the main event hall at opposite corners; therefore, two of five teams were stationed there, enabling them to work together and avoid interfering with each other in both technical and organizational matters.
The other three teams were located in adjacent rooms, that also enabled a non-interfering environment i.e. in terms of non-overlapping tracking systems. 
At the main hall a variety of different solutions were at teams' disposal: from affordable cardboard, through standalone mobile devices like Oculus Go and Quest, up to top-notch Varjo VR-1 with UHD display. Nevertheless, during the main part of the hackathon each team was provided an with individual comparable VR development set in non-blocking configuration, that comprised of wired HMD (HTC Pro or Oculus Rift class, with standard base stations and controllers
) hooked up to custom workstation based on the latest AMD Ryzen 7 class CPU platform coupled with Nvidia GeForce RTX 2080 class GPUs.

\subsubsection{Experts' Support}

The teams were supported by experts from the participating institutions and universities who were there on site to provide mentoring and guidance. The guidance ranged from technical expertise on Unity or Unreal, through psychology, teaching and learning methodology and art guidance on the key principles of Bauhaus. The teams participated in initial training on these topics, but later they could consult the experts as needed.

\begin{figure}[tb]
 \centering 
 \includegraphics[width=\columnwidth]{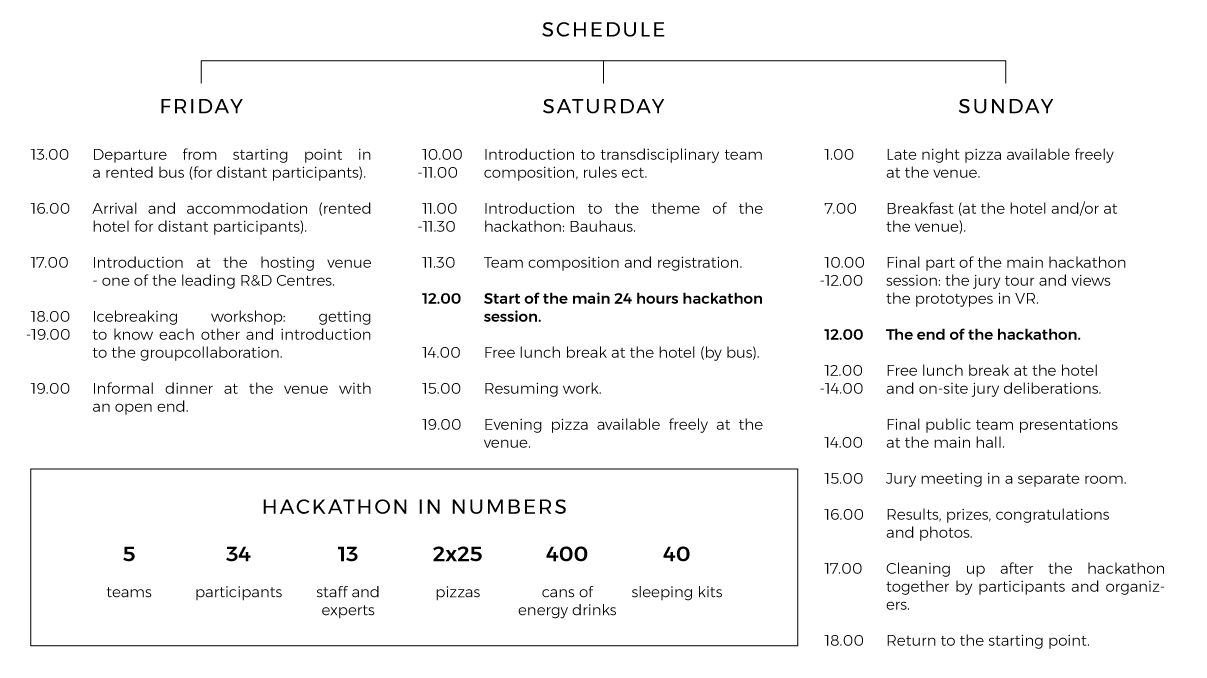}
 \caption{Overview of the hackathon events.}
  \Description{Overview of the hackathon events as a schedule listing activities as well as key figures related to the execution of the hackathon.}
 \label{fig:overview}
\end{figure}

\subsection{Collaboration and Outputs}

\subsubsection{Team formation and composition}

An important part of enabling better team collaboration was a one-hour integration workshop. During the workshop, warm-up and ice-breaking were merged with practical exercises providing basic psychoeducation and to participants and enhancing their communication skills in order to empower interdisciplinary cooperation. This main aim was achieved through: 
\begin{enumerate}
    \item underlining the importance of communication, including the adjustment of the message to the audience coming from a different scientific background;  \cite{callaos_horne_2013}
    \item working out an appropriate amount of message content enhancing memorization and taking into account cognitive functions efficiency and environmental distractors; \cite{whosaidhwat}
    \item group roles while solving a complex problem requiring creative thinking and \cite{problemsolving}
    \item enhancing both group and individual creativity. \cite{creativity}
\end{enumerate} 
Two conducted exercises addressed all four points mentioned above. \footnote{This, and the whole event, is well-illustrated in our hackathon summary video, which is available on YouTube, at the XRLab PJAIT channel here: https://youtu.be/uZhgHDv2xPk} In the first exercise, addressing 1 and 2, participants had to choose a Dixit card illustrating their expectations about the hackathon and to describe it briefly on a post-it. With the principle of speed dating, they had one minute to present the card to another participant, then exchange each other's cards and move on to the next participant. At the end of the exercise, each participant described the last received card and the group compared the description with the original one written down on the post-it. In the second exercise, addressing 3 and 4, participants were divided into 6 groups, each group had to solve a complex tangram-like puzzle requiring creative thinking (elements had to be turned inside out in order to complete the puzzle). The exercises were discussed from the perspective of communication, memory and executive functioning, and the sum-up was focused on group roles and individual and group creativity. 
 
The next day there was an opportunity to rethink the strategy of team formation and composition during the second part of empowerment workshops in the morning, which also brought back themes discussed during the kickoff meeting two weeks prior. After the introduction of the outline of the competition and foundations of effective transdiciplinary teamwork, participants had another opportunity to get to know each other before final team formation. Every participant had to present publicly their own motivation, expectations, background and expertise to provide equal distribution of information among participants. They also put on the sticky notes their names with key information, i.e. skills or area of expertise, important from the viewpoint of transdisciplinary team composition. The groups of participants included:
software developers (with specialization: Unity/Unreal/other), designers (with optional details on modeling/animation/other), psychology and language. This procedure facilitated the process of finalizing team formation and the following registration session. As a result five transdiscyplinary teams were composed, each of which consisted of members of various expert-fields: software development (Unity or Unreal), design, psychology and linguistics.

\begin{figure}[tb]
 \centering 
 \includegraphics[width=\columnwidth]{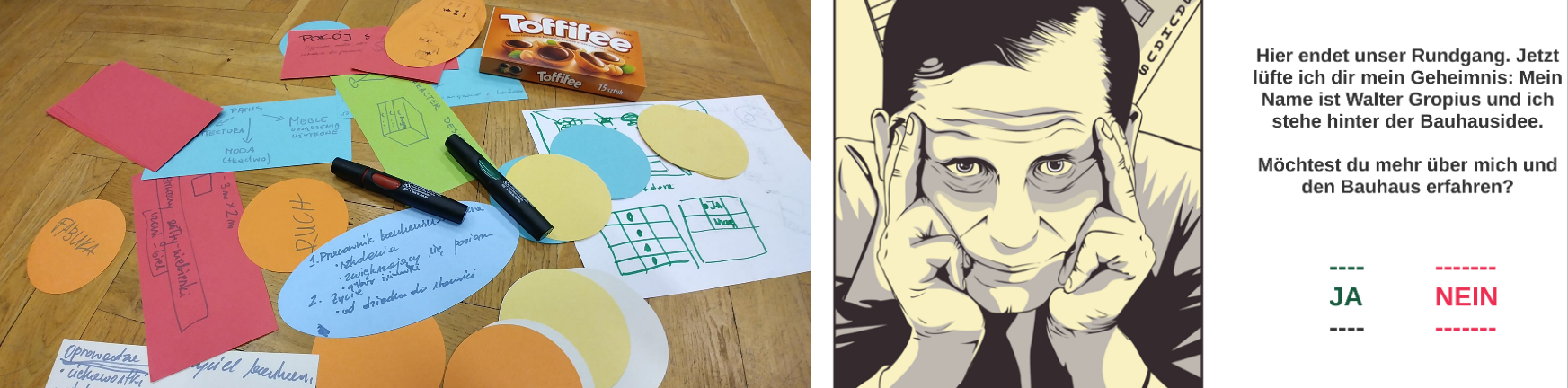}
 \caption{Left: Some teams started with paper prototyping and creating mash-ups to come up with the idea for their VR app. This allowed all team members to contribute equally from the start. [Team 4]. Right: Interaction concept from the final presentation [Team 2]}
  \Description{Left - example of a mash-up - : Some teams started with paper prototyping and creating mash-ups to come up with the idea for their VR app. This allowed all team members to contribute equally from the start. [Team 4]. Right - presentation slide showing the conceptual work - : Interaction concept from the final presentation [Team 2]}
 \label{fig:conceptpres}
\end{figure}

\subsubsection{Judging Criteria}

The Jury consisted of 13 experts from various institutions, including the Goethe Institute, technical universities and innovation-driven NGOs, each of whom had 9 points to award, 3 in each category: \textbf{innovation, functionality, feasibility}. In total, 117 points were awarded. The assessment relied on jury members' expert knowledge and experience applied to those three crucial dimensions; special attention was given to the functions required in the process of learning languages and project viability. Each team was granted freedom in regard to their willingness to further work on their project, but the important criterion was whether the project could be developed from the hackathon hi-fidelity prototype into a fully-fledged app with the support of the organizers. As a result, we ended up with a comprehensive, multidimensional assessment of the projects. If the result did not satisfy the Goethe Institute representatives they had the right to change the ranking of the projects (which was not used).

\subsubsection{Projects}
In this subsection we present the projects in order of teams' registration number.

\begin{itemize}
\item \textbf{Team 1 - Follow the Path}
\end{itemize}

The team created a labyrinth with multiple rooms where the users could engage with various language learning games at an easy level; 
to proceed to the next room you had to receive a key after solving the puzzle correctly. The color scheme followed the key colours of Bauhaus and the aesthetics were aligned with its principles. The project was the most robust and functional, and it had great potential for further development. The team got \textbf{second place}, with 28 points, 8 for innovation, 11 for functionality and 9 for feasibility.
\begin{figure}[tb]
 \centering 
 \includegraphics[width=\columnwidth]{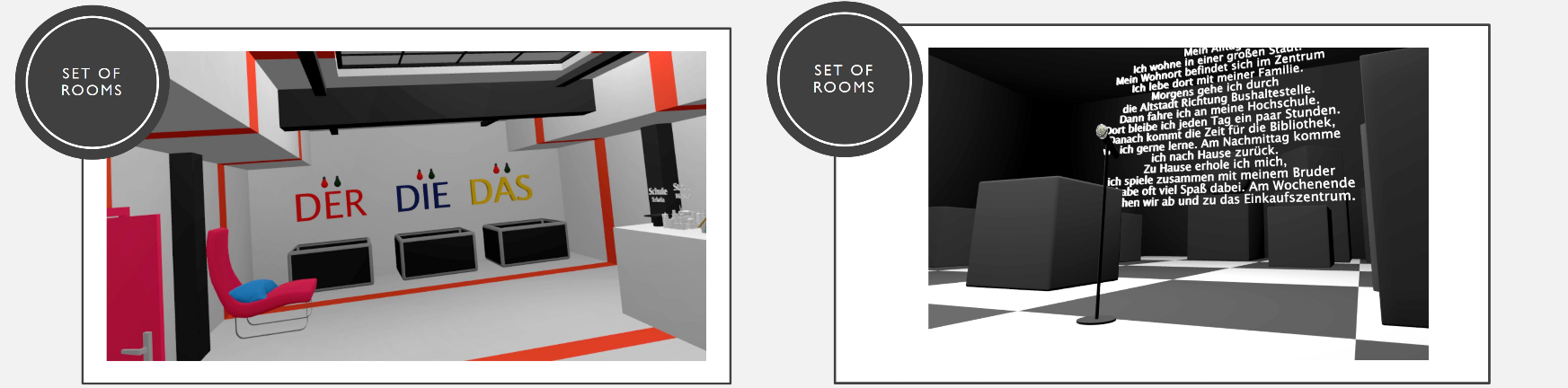}
 \caption{Left: In this room the users could throw items into the right basket depending on the words they heard to indicate which article (der, die or das) the word takes. [Team 1] Right: One of the rooms allowing users to test their reading skills, the idea was to use voice recognition using Azure Speech to Text, Amazon Transcribe and Amazon Comprehend to evaluate the user's performance. [Team 1] }
 \Description{Left - a screenshot of a room from the VR app of [Team 1] Right: another room with text to read shown in the centre - [Team 1] }
 \label{fig:team1a}
\end{figure}

\begin{itemize}
\item \textbf{Team 2 - Travel}
\end{itemize}

The users could walk around in Street View, and the project assumed that they could stop by any hotspots or attractions, such as museums and restaurants and interact with them. The project was at an early concept stage, but it received recognition due to the idea of using the language in a natural setting. 
However, it was not very functional, as only some UI elements were present in VR together with Google Street View that one could explore, and there was not much interactivity. The team got \textbf{third place}, with 25 points, 8 for innovation, 6 for functionality and 11 for feasibility.

\begin{itemize}
\item \textbf{Team 3 - Escape German}
\end{itemize}

The project consisted of an Escape Room in VR where the users were to solve puzzles in German to find clues that would help them to leave the room. However, the team did not have many ideas of how to create German language puzzles that could help learning and did not implement them in VR. The team got fifth place, with 3 points, 2 for innovation, 1 for functionality and 0 for feasibility.

\begin{itemize}
\item \textbf{Team 4 - Motion Craft}
\end{itemize}

The user appears in an environment with Bauhaus-style furniture outlines, and for the furniture to appear they have to put them together in a workshop area following audio instructions in German. The team prototyped one room where the users could put together a piece of furniture. The idea to scale it was to add more workshops with different furniture and objects, so the whole project seemed to be easily expandable based on a concept of subsequent rooms.
The project was well-received in terms of the linguistic element because of the learning-by-doing factor. Overall the team got \textbf{first place}, with 39 points, 15 for innovation, 16 for functionality and 8 for feasibility.

\begin{figure}[tb]
 \centering 
 \includegraphics[width=\columnwidth]{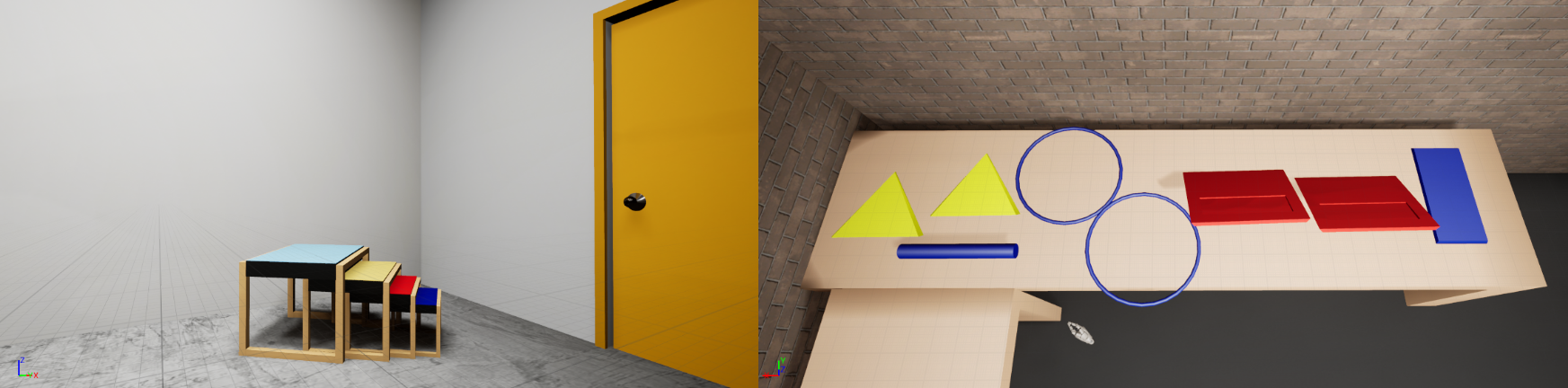}
 \caption{Left: The users had to put furniture together for it to appear in their environment.  Right: The workshop area where the users would put the furniture together after hearing instructions in German. [Team 4] }
 \label{fig:team4b}
\end{figure}

\begin{itemize}
\item \textbf{Team 5 - Bauhaus Jazz}
\end{itemize}

The user entered a Bauhaus-style room with key information about the Bauhaus movement on a wall. The user was to throw a ball, a cube or a cone at different things in the room depending on the article they take (der, die or das). If the user threw correctly they got a point so the game was competitive as it allowed for performance comparisons. The project was well-made and quite finished, however, the linguistic element was not prominent and the team member responsible for the language part had very little to do, and finally withdrew from teamwork due to this. The team got fourth place, with 22 points, 6 for innovation, 5 for functionality and 11 for feasibility.
\subsubsection{Recognition and Rewards}

The winning projects continued to be developed under the tutorage of the experts taking part in the hackathon. 
During the first quarter of 2020, the teams started intense work on porting the whole solution into Unity, predominantly having in mind further expansion of the project by other teams and the capability of coupling the project with the psychophysiological framework for further in-depth research in terms of user interaction, reception and learning efficiency compared to other methods. Ultimately, the winners were to participate in a special VR summit and training event at Goethe-Institut Berlin and show off their projects; however, the event was postponed due to the COVID-19 pandemic.

\section{Discussion and Lessons Learned}

\subsection{Preparation}

The organizers should have provided more comprehensive support concerning modern language teaching and learning methodology, especially that Goethe and DAAD were available to aid the process, and that, they as the clients/partners, did not want a traditional approach to language learning to appear in the developed VR applications. In retrospect, the hackathon participants specializing in teaching methodology and language learning would also have benefited from a workshop on this topic organized by Goethe and DAAD, as they did not share the same views on what constitutes modern language-learning methodology as the judges/clients. An example of a successful theme introduction is the presentation of Bauhaus concepts of art and design. A few general slides and broad strokes on the history of this art school and its principles together with a link to a website that collected information about Bauhaus were sufficient to see some references to Bauhaus in each project.

\subsection{Execution}

Useful observations and insights have been gained over the execution of the case study. First, initially there were 6 teams, but the last team quickly joined team number 4, forming a larger team that ultimately won the competition. Another important fact relates to the team diversity and composition. Most of the teams followed the guidelines conveyed during workshops and empowerment sessions and formed very diverse teams in terms of expertise as well as gender balance. In contrast, one of the teams (5) remained unbalanced till the end. Although its members were highly skilled in terms of designing and coding, they lacked both: gender balance and expertise coverage, with only one expert (although in a double role: as a psychologist and linguist).

\subsection{Impact and Afterthoughts}

\subsubsection{Integration and Empowerment}
We have noticed marked benefits of working in transdisciplinary environments for the members of each team, such as quick knowledge acquisition on unfamiliar topics, learning to work with ambiguity and to collaborate with representatives of different disciplines. However, despite the requirement that the team includes representatives of each of the four expert areas collaborating and exchanging expertise, some team members seemed to engage to a lesser extent. Having that in mind in the subsequent follow-up steps after the hackathon we put more emphasis on strengthening this area by organizing special empowerment sessions with experts from the Goethe Institute and in-depth language workshops.
Moreover, the fact that during this hackathon paper prototyping stage was very short and consisted of brainstorming during idea development without much actual drawing on paper aided in making the applications more feasible, and it made it more motivating for the participants to continue work on them after the hackathon. This is in line with the assumptions of project-based learning, where practical application of knowledge speeds up the learning process. It is very important also that there was always a sight of the end goal: the presentation of the apps in Berlin, so that the participants knew that their work will not be shelved.

\subsubsection{Rapid Design Cycle for VR and Continuing Projects}

The format of the hackathon necessitates making quick design decisions and going from ideas, through prototypes towards working solutions at a very fast pace. On the other hand, without proper empowerment some participants of the team may miss their chance to veto an idea, as they have little prior experience with the XR solutions' continuum. However, because the solutions in VR may be prototyped quite quickly with ready assets, even the team members who had little prior experience can start to see what the end goal is and help with shaping it as VR testers. 

Overall, we believe that prototyping in VR, rather than on paper/with clay may be more effective. First, interaction in VR is very intuitive and the users/testers can quickly grasp the concepts 
which they may have otherwise have trouble imagining. Jumping straight to prototyping in VR, without creating outside mock-ups which function on a metaphorical level, helps people who do not share the same understanding of the metaphor contribute valuable insights early.

Second, prototyping applications which are usable encourages the teams to continue their work after the hackathon.
This, of course, necessitates that in every VR hackathon team there is at least one VR developer who can create such prototypes.

\subsubsection{Our Recommendations}

\begin{enumerate}
\item \textbf{Team composition}
We recommend that in each team prototyping for VR there is at least one developer who can make a functional prototype and who can be engaged enough to be able to take this prototype further after the hackathon is done.
\item \textbf{Integration}
The integration aspect was critical for the participants to gain an understanding of their goals for the hackathon, and to learn to work as a group. Especially solving the special tangram-like puzzles together helped create a sense of camaraderie and aided in kicking off work the next day.
\item \textbf{Empowerment}
This element is critical, as without it some team members may be excluded or overlooked.
\begin{itemize}
\item \textbf{VR empowerment}
For participants who have little experience with VR it is important to show them the range of this technology, the possibilities and the way one interacts with the environment. In our case, this was done in several ways, also including trying out the Beat Saber game in VR.

\item \textbf{Sufficient need explanation}
The hackathon organizers ought to provide the clients with expert assistance in the area of participatory design and requirement analysis to assist in explaining their needs in detail, as they may not realize the extent of knowledge (or lack thereof) of the participants.

\item \textbf{Presentation team members' expertise}
It is important to place the participants most strongly related to the theme at the centre. Especially team diversity and team members voice equality are two important fundamentals. From our experience, all of the participants should be introduced as experts in their respective fields who have specific and crucial knowledge on the topic at hand.
\end{itemize}
\item \textbf{Idea generation}
The importance of the idea generation phase should be underlined, and the participants should be provided with post-it cards and other supplies and encouraged to use them. This is especially crucial as we recommend the paper prototyping phase to be shortened considerably in favour of creating working prototypes in VR.
\item \textbf{Expert feedback} 
The experts have to be instructed to attempt to help each team equally. Some of the experts were disproportionally used by some teams, and later they, because of familiarity, felt the need to check on them more often, which only widened the gap. This was especially visible with language and technology.
\item \textbf{Prototyping}
We recommend actual prototyping to be done already in VR and to encourage the teams to build working prototypes, as this helps with both testing in a participatory manner during the hackathon, and later the actual implementation.
\item \textbf{Follow-up}
Knowing what will happen to the prototypes later
helps the team remain motivated. Hackathons, in general, suffer from low implementation levels, but making working prototypes is a great way to not only encourage further development, but also garner interest from others after the hackathon ends, including investors.
\end{enumerate}

\section{Conclusions}

The results of this study cast light onto the recommended cycle of design thinking and customer-centered design in VR, showcasing the need to create working prototypes in the low-fidelity development stage to enable stakeholders to experience, and thus, imagine the end product. The better the early immersion, the more valid the insights all team members can produce. They also showcase the educational benefits of working in transdisciplinary environments, such as quick knowledge acquisition, working with ambiguity and learning to collaborate with representatives of different disciplines. However, the non-technical members of teams need to be encouraged to be more vocal and active.
We are also looking forward to testing more participatory and hands-on ways of designing for VR, using theatre techniques and improvisation, which would allow to quickly test early prototypes by acting out scripts rather than building physical or virtual mock-ups. 

This study, combined with our previous work on human factors in rapid software development and co-design, including hackathon dynamics, allowed us to formulate recommendations for organizing content creation VR hackathons for specific purposes. We also provided guidelines on how to prepare the participants to work in rapid prototyping VR environments and benefit from such experiences in the long term. One key suggestion is to organize participatory design workshops with the client who has a specific purpose before the hackathon, to work on the concept of the hackathon and specify exact requirements prior to its execution and empower all of the participants to make the best use of their skills.

\section{Acknowledgments}
We would like to thank all the members of our Living Lab, those affiliated with the Kobo Association who participated in this study and all transdisciplinary experts involved with the HASE research initiative (Human Aspects in Software and System Engineering). We also would like to thank all the students and partners involved in this research activity and events, especially Goethe Institut. This cooperation could not happen without the direct involvement of our late professor Krzysztof Marasek, who enabled the cooperation between the research group of our Living Lab and Goethe Institut. We treat the work reported in this article and our further research as a special tribute to our late professor which we think commemorates him better than any kind of monument he deserves.

\bibliographystyle{ACM-Reference-Format}
\bibliography{bibliography}

\end{document}